\documentclass[prd,groupedaddress,nofootinbib,showpacs]
{revtex4}

\usepackage{latexsym}
\usepackage{amssymb}
\usepackage{graphicx}
\usepackage{dcolumn}
\usepackage{bm}
\input{epsf}
\begin{document}

\title{PLANE GRAVITATIONAL RADIATION FROM NEUTRINOS SOURCE WITH KALB-RAMOND COUPLING}

\author{C. N. FERREIRA}


\affiliation{N\'ucleo de Estudos em F\'{\i}sica, Centro Federal de Educa\c{c}\~ao Tecnol\'ogica de Campos \\
Rua Dr. Siqueira, 273, Campos dos Goytacazes, Rio de Janeiro, Brazil, CEP 28030-130\\
crisnfer@cefetcampos.br}

\author{J. A. HELAY\"EL-NETO}


\affiliation{Centro Brasileiro de Pesquisas F\'{\i}sicas, \\ Rua Dr. Xavier Sigaud 150,
Urca, Rio de Janeiro, Brazil, CEP 22290-180 \\
helayel@cbpf.br}

\author{N. A. TOMIMURA}

\affiliation{Centro Brasileiro de Pesquisas F\'{\i}sicas, Rua Dr. Xavier Sigaud 150,
Urca,\\ Rio de Janeiro, Brazil, CEP 22290-180 \\
and \\ Instituto de F\'{\i}sica, Universidade Federal Fluminense, \\
Av. Litor\^anea s/n, Boa Viagem, 24210-340, Niter\'oi, RJ, Brazil \\
nazt@if.uff.br}

\date{\today}



\begin{abstract}
In this work, we propose a model based on a non-minimal coupling of neutrinos to a Kalb-Ramond 
field. The latter is taken as a possible source for gravitational radiation. 
As an immediate illustration of this system, we have studied the case where 
gravitational (plane) wave solutions behave as damped harmonic oscillators.

\end{abstract}



\maketitle

\section{Introduction}

Neutrino Physic has had a remarkable progress in the last two decades.\cite{1}  However, a satisfactory picture of the neutrino properties are not fully-understood yet. Neutrinos are subatomic particles generated in weak interaction processes that have an extremely small scattering cross section. In their interactions, neutrinos exhibit parity violation. In the Standard Model (SM), the matter interaction occours through their left-handed chiral projection and they are massless particles like photons. The solar neutrino problem has driven a great deal of interest in the neutrino oscillations and neutrino masses, which are  contemplated and discussed in the so-called Physics Beyond the Standard Model.
Recently, neutrino oscillations have been analyzed experimentally by Superkamiokande\cite{1} and the Sudbury Neutrino Observatory (SNO)\cite{2} demonstrating 
that neutrinos can transform from one species to
another while propagating through space.\cite{3}
In the Universe, there is an enourmous 
number of neutrinos, for instance, there are neutrinos emitted during the onset of gravitational collapse to explosion.\cite{Kotake:2005zn}
At the moment of explosion, most of the binding energy of the core is released as neutrinos. The core-collapse supernovae have been supposed  to be one of the most plausible sources for gravitational waves. The study of both phenomena can help us to understand the core colapse supernovae, the properties of neutrino and gravitational waves. Some autors study the neutrinos related with gravitational waves considering an imperfect fluid.\cite{Weinberg:2003ur,Dicus:2005rh,Miao:2007cw,Wang:2008fg}
Our idea is that, during the process of supernovae destabilization,  neutrinos interact non-minimally with the Kalb-Ramond fluid that can be interpreted as related with the shock waves. The neutrinos resulting from this process are in a burst form and generate gravitational waves.

\section{The Possibles Fermionic Sources to Kalb-Ramond Gravitational Coupling}

In this section, we analyze the Kalb-Ramond interaction Lagrangean that can describe the neutrinos in our approach. 
The effective Lagrangean that describes the coherent neutrino scattering, without mass, in a dense medium has the form
\begin{equation}
{\cal L}_{Int} = \sum_a \alpha_a(X_a) (\bar \nu \Gamma^a \nu), \label{lagrangian1}
\end{equation}
where $\alpha_a$ is the coupling constant that depends on the a-type of the bilinear term.
Among the whole class of $ \Gamma^a$'s
$\Gamma^a=  I, \gamma_5, \gamma^{\mu},\gamma^{\mu}\gamma_5 $, $\sigma^{\mu \nu}$ with $\sigma^{\mu \nu} = {i \over 2}[\gamma^{\mu}, \gamma^{\nu}] $,
the scalar (I) and pseudoscalar ( $\gamma_5 $ ) currents are suppressed for highly relativistic neutrinos.

In the Lagrangian (\ref{lagrangian1}), the notation $X_a $ are the complex scalar $(X_S)$, pseudoscalar $(X_p)$, vector $(X_{V\mu})$, axial-vector $(X_{A\mu})$, and tensor $(Z_{T\mu \nu})$ characterizing the background. 

In the highly relativistic limit of  neutrino scattering, the scalar and pseudo-scalar terms can be effectively omitted. If, in addition, we assume that the tensor term is negligeable (since it is merely a weak spin interaction), the interaction lagrangian (\ref{lagrangian1}) reduces to 
\begin{equation}
{\cal L}_{Int} = -\eta \, \bar \nu \gamma^{\mu} (1- \gamma_5) \nu \, X_{\mu}.\label{leftinteration}
\end{equation}
In our problem, we consider a new interpretation by considering the gravitational waves induced by the interaction among neutrinos and the Kalb-Ramond fluid with $X_{\mu} = \tilde G_{\mu}$, where $\tilde G_{\mu} = \frac{1}{3 !}\epsilon_{\mu \nu \alpha \beta} G^{\nu \alpha \beta}$,
where $\tilde G_{\mu}$ is the dual fieldstrength, given by
$G_{\mu \nu \kappa} = \partial_{\mu}B_{\nu \kappa} + \partial_{\nu}B_{\kappa \mu} + \partial_{\kappa}B_{\mu \nu}$.
$B_{\mu \nu}$ is the skew-symmetric 2-form gauge potential referred to as the Kalb-Ramond field.

The Bianchi identities for the Kalb-Ramond field read as  $ \partial_{\mu} \tilde 
G^{\mu}=0$.
In this work, let us describe the neutrinos in a curved space-time; in this case, the covariant derivative changes to $\nabla_{\mu} $ and the Bianchi identity becomes $
\nabla_{\mu}\tilde G^{\mu}= \partial_{\mu}G^{\mu} + \gamma_{\nu \mu}^{\mu} \tilde G^{\nu} =0 $; it is easily shown that this equation satisfies the energy conservation equation,  $T^{\mu \nu}_{\, \, \, ; \mu} =0$, and this fact is compatible with gravitational couplings.

\section{Einstein Equations to Plane Gravitational Waves}

In this section, we apply the model described above to find an exact solution for plane gravitaional  waves. 
Our metric in null coordinates reads as below:
\begin{eqnarray}
d^2s &=& L^2 \Big(e^{2 \beta} dx^2 + e^{-2\beta}dy^2\Big) + dz^2 - dt^2\nonumber \\ 
&= & L^2 \Big( e^{2 \beta} dx^2 + e^{-2 \beta}dy^2\Big) - du dv \label{metric},
\end{eqnarray}
where $u = t-z$, $v=t+z $, $L=L(u) $ and $\beta = \beta(u) $.
In the null coordenate system $ u, v, y, z $, the only  component of the Ricci tensor that does not vanish identically is $
R_{uu}  =  - L^{-1}\Big(L'' + \beta'^2 L\Big)$.
Einstein' s equation in this system is given by
\begin{equation}
- L^{-1}\Big(L'' + \beta'^2 L\Big)= {\kappa \over 2} T_{uu}, \label{motion1}
\end{equation}
where $ T_{uu}$ is the energy-tensor in light-cone frame. In our treatment, we perform the calculation of the contribiution of the neutrino-Kalb-Ramond coupling to the energy-momentum tensor in the (t, z, x, y)-frame and, since  the radiation propeties are better analysed in light-cone frame. We transform the results to this frame and, using the definition  $
T_{\mu \nu} = {2 \over \sqrt{-g}}{\delta S \over \delta g^{\mu \nu}} $, we have the relations  $ T_{uu} = -T_{tz}$ , $T_{uu} = - \eta \Big(\bar \psi \gamma_t\tilde G_z \psi + \bar \psi \gamma_z\tilde G_t \psi\Big) $. 
We consider the positive helicity; in this case, if we adopt the Weyl representation of the Dirac $ \gamma^{\mu}$- matrices,  the spinor reduces to
$
\psi = \left(\begin{array}{ll}
\lambda &  \\
0 & 
\end{array} \right)$. Using the constraints, we have
\begin{eqnarray}
T_{uu}  &=& -2 \eta  |\lambda |^2 \tilde G_t,  \label{tensor1}
\end{eqnarray}
where $ \tilde G_t = \tilde G_u$. This result is compatible with the physical interpretation that the gravitational wave are generated by imperfect fluid.\cite{Weinberg:2003ur} 

In our approach, the energy-momentum 
tensor exhibit diagonal and off-diagonal  componets. When the off-diagonal part is zero, $T_{ij} =0$, the only contribution of the 
radiation is the plane wave in the vacuum,    
which implies $T_{uu} =0$. This does not give an information about the source.

The configuration of the Kalb-Ramond field that is compatible with the plane 
wave metric (\ref{metric}) and respects the energy-momentum tensor conservation. From
$
\partial_{\mu} \tilde G^{\mu}= 0$, it may be written that  
$\tilde G_{\mu} = \partial_{\mu} \phi $.
The possible solution that satisfies this condition is 
$\phi =  {\rm ln} L$, then the Einstein equation (\ref{motion1}) for $ \beta'^2=\omega_0^2$ reduces to
\begin{equation}
L'' + 2\gamma L' + \omega_0^2 L =0,
\end{equation}
where $\gamma = {1 \over 2} \kappa \eta |\lambda |^2$. 
The solution to this equation is given by $
L = A\sin(\omega  \, u)e^{-\gamma u} $
where $\omega^2 = \gamma^2-\omega_0^2$.  
That give us the energy loss with u as $
E = {\omega^2 A^2 \over 2} e^{-2 \gamma u} $, when the damping is small $(\gamma << w_0)$. 
We can interpretat the energy lost, $ E$,  the mass variation in the volume.

\section{Conclusion}
In this work, we have shown that the Kalb-Ramond coupling to neutrinos  can  be a viable source  of gravitational radiation. The exact plane wave metric gives us the usual topological equation for the dual of the Kalb-Ramond field-strength that can be written as the gradient of a scalar.
In this case, when this scalar field is logarithmic in the length scale  of the metric, we can find an damped harmonic oscillator. When the damping is small, $\gamma << w_0$ , the energy related is
approximately,  $ E_{\nu} = {1 \over 2} \omega^2 A^2 e^{-2\gamma u}$.
Thus, the energy falls off exponentially at twice the frequency of which the
amplitude decays.  
In future works, we have in mind to propose a numerical situation that can be compatible with the supernovae 
explosions and study the problem of the gravitational wave generated by neutrinos oscillations.

\section*{Acknowledgments}

CNF and JAHN would like to express their gratitude to CNPq-Brasil and NAT to 
FAPERJ for the invaluable financial support.

\end{document}